\documentclass[review,12pt]{elsarticle}
   
\usepackage[cmex10]{amsmath}
\usepackage{amssymb}
\usepackage{amsfonts}
\usepackage{array}
\usepackage{mathrsfs}
\usepackage{comment}
\usepackage{graphicx}
\usepackage{rotating}
\usepackage{subfigure}
\usepackage{bm}
\usepackage{url}
\usepackage{bm,amssymb,amsthm}
\usepackage{algorithm}
\usepackage{algorithmic}
\usepackage{multirow}
\usepackage{standalone}
\usepackage{blindtext}
\usepackage{float}
\usepackage{hyperref}
\usepackage{bm}
\usepackage{makecell}
\usepackage{array}
\usepackage{subscript}
\usepackage{textcomp}
\usepackage{caption}
\usepackage{setspace}
\usepackage{ctable}

\newcommand{\ev}[1]{\mathbb{E}[{#1}]}
\newcommand{\cov}[1]{\bm{\mathcal{D}}[{#1}]}
\newcommand{\bth}{\bm{\theta}}
\newcommand{\argmin}{\operatornamewithlimits{argmin}}
\newcommand{\pd}{\no\hspace{2em}{\itshape Proof: }} 
\newcommand{\pds}{\no\hspace{2em}{\itshape Sketch of proof: }}

\mathchardef\mhyphen="2D

\newcommand{\gr}[1]{\bm{#1}}

\newcommand{\grl}[1]{\textbf{#1}}

\newcommand{\q}{\grl{q}}
\newcommand{\Q}{\gr{Q}}
\newcommand{\Y}{\gr{Y}}

\newcommand{\A}{\gr{A}}
\newcommand{\B}{\gr{B}}

\newcommand{\I}{\gr{I}}

\newcommand{\K}{\gr{K}}
\newcommand{\C}{\gr{C}}
\newcommand{\D}{\gr{D}}
\newcommand{\Z}{\gr{Z}}

\newcommand{\W}{\gr{W}}
\newcommand{\R}{\gr{R}}
\newcommand{\N}{\gr{N}}
\renewcommand{\H}{\gr{H}}
\renewcommand{\S}{\gr{S}}
\newcommand{\G}{\gr{G}}
\newcommand{\T}{\gr{T}}

\newcommand{\X}{\gr{X}}

\newcommand{\U}{\gr{U}}

\newcommand{\V}{\gr{V}}

\newcommand{\Pre}{\gr{P}}

\newcommand{\n}{\grl{n}}

\newcommand{\x}{\grl{x}}

\newcommand{\y}{\grl{y}}

\newcommand{\sgens}[1]{\mathbb{R}^{#1}}
\newcommand{\sgen}[2]{\mathbb{R}^{#1\times #2}}

\newcommand{\Si}{\gr{\Sigma}}
\newcommand{\La}{\gr{\Lambda}}

\newcommand{\no}{\noindent}

\newtheorem{theorem}{Theorem}
\newtheorem{lemma}{Lemma}

\newtheorem{proposition}{Proposition}

\newtheorem{definition}{Definition}
\newtheorem{remark}{Remark}
\newtheorem{fact}{Fact}

\journal{Biomedical Signal Processing and Control}

\begin{document}

\begin{frontmatter}
  
\title{Localization of Brain Activity from EEG/MEG Using MV-PURE Framework}

  \author[TP,TP2]{Tomasz Piotrowski\corref{cor1}}
  \ead{tpiotrowski@is.umk.pl}
  \cortext[cor1]{Corresponding author.}
  \fntext[fn1]{Published: \url{https://doi.org/10.1016/j.bspc.2020.102243}}
  \address[TP]{Department of Informatics,\\
    Faculty of Physics, Astronomy and Informatics,\\
    Nicolaus Copernicus University,\\
    Grudziadzka 5, 87-100 Torun, Poland}
  \address[TP2]{Centre for Modern Interdisciplinary Technologies,\\
    Nicolaus Copernicus University,\\
    Wilenska 4, 87-100 Torun, Poland}
  \address[AM]{Behavioral and Cognitive Neuroscience Institute,\\
    Simon Fraser University,\\
    8888 University Ave. Burnaby BC, V5A 1S6, Canada}

  \author[TP,TP2]{Jan Nikadon}
  \author[AM]{Alexander Moiseev}

\begin{abstract}
We consider the problem of localization of sources of brain electrical activity from electroencephalographic (EEG) and magnetoencephalographic (MEG) measurements using spatial filtering techniques. We propose novel reduced-rank activity indices based on the \emph{minimum-variance pseudo-unbiased reduced-rank estimation (MV-PURE)} framework. The main results of this paper establish the key unbiasedness property of the proposed indices and their higher spatial resolution compared with full-rank indices in challenging task of localizing closely positioned and possibly highly correlated sources, especially in low signal-to-noise regime. Numerical examples are provided to illustrate the practical applicability of the proposed activity indices using both simulated and real data. 
\end{abstract}

\begin{keyword}
EEG, MEG, spatial filters, neural activity indices
\end{keyword}

\end{frontmatter}

\section{Introduction}
Electroencephalography (EEG) and magnetoencephalography (MEG) enable measuring electromagnetic fields generated by synchronous post-synaptic electric currents in populations of neurons, using an array of sensors positioned outside the brain \cite{Hamalainen1993, Mosher1999}. Adaptive spatial filters (beamformers) have been used in array signal processing since the seminal paper by Frost \cite{Frost1972}. In EEG and MEG, beamforming has been used for source localization and reconstruction of electrical activity of sources (see \cite{Baillet2001, Greenblatt2005} for a review). One of the widely used spatial filters is the linearly constrained minimum-variance (LCMV) beamformer \cite{VanVeen1997, Gross2001, Sekihara2001, Sekihara2008, Moiseev2011, Diwakar2011, Moiseev2015, Kuznetsova2019, Chrapka2019}. The LCMV beamformer is implemented in virtually all software enabling EEG/MEG source analysis \cite{FieldTrip2011, Brainstorm2011, MNE2014, Jaiswal2019} and continues to find research use in EEG/MEG community, see, e.g., \cite{Keitel2016, Siems2016, Nunes2020}.

In this paper we focus on source localization from EEG/MEG measurements. The beamformer-based approach to this problem is based on the pioneering idea of an activity index (localizer function) named \emph{neural activity index (NAI)} introduced in \cite{VanVeen1997}. The NAI is a function of source location and is defined as a certain power ratio of the output of the LCMV filter. Its main assumption is that its maxima coincide with locations of active sources. Indeed, several other activity indices based on LCMV beamformer have been introduced since \cite{VanVeen1997}, see, for example, \cite{Dalal2006, Popescu2008, Quraan2010}. However, neither $NAI$, nor the newer activity indices, have been proved to be spatially unbiased, i.e., that they achieve their global maxima if evaluated at locations of sources responsible for observed data. In other words, their main assumption relies in each case on a certain heuristic approach. Only recently, Moiseev et al. \cite{Moiseev2011} has proved unbiased property of some traditional single source activity indices (localizers) based on the LCMV beamformer in a general case, and introduced their generalizations for multi-source spatial filters known as Multiple Constrained Minimum Variance (MCMV) beamformers. Remarkably, the unbiased property theoretically holds for any signal-to-noise ratio and any level of correlation between the sources. Yet, in practice, these factors strongly affect spatial resolution of activity indices.\footnote{Spatial resolution is defined in Definition \ref{spatial_res_df} in Section \ref{preli}.} Moreover, it has been demonstrated in \cite{VanVeen1997, Sekihara2001, Sekihara2008, Diwakar2011} and references therein that the single source LCMV solutions may perform sufficiently well only if certain conditions are satisfied, such as zero or low correlations between the sources, high enough signal-to-noise ratios (SNRs) and sufficiently large spatial separation of sources. Thus, it would be desirable to derive activity indices which do not inherit the drawbacks of the single source LCMV beamformer and offer possibly higher spatial resolution.

In this paper we introduce multi-source activity indices built on the minimum-variance pseudo-unbiased reduced-rank (MV-PURE) estimation framework \cite{Yamada2006,Piotrowski2008,Piotrowski2009,Piotrowski2014b,Piotrowski2018s}. We prove that these indices are also unbiased and show that they provide higher spatial resolution than MCMV localizers in ill-conditioned settings, such as when the sources are closely positioned and/or exhibit highly correlated activity, or when the SNR is poor - that is, in situations often encountered while analyzing neurophysiological data.

The proposed activity indices are natural generalizations of the MCMV localizers proposed in \cite{Moiseev2011}, just as the MV-PURE framework generalizes the MCMV approach to the reduced-rank case. Specifically the activity indices proposed in this paper are \textit{multi-source}, meaning that they are functions of \textit{all} sources' positions and orientations (in a general case). This fact renders an exhaustive brute force search for source parameters infeasible, as that would amount to testing exponentially growing number of possible combinations of locations and orientations of all simultaneously active sources. Thus, for the proposed activity indices we employed an iterative sequential search procedure introduced in \cite{Moiseev2011} which allows finding the sources with reasonable accuracy and computational effort.

The paper is organized as follows. In Section \ref{preli} we introduce the notation, the EEG/MEG forward model, and the definitions of an unbiased activity index and its spatial resolution. We also provide definitions of full-rank activity indices introduced in \cite{Moiseev2011} along with their generalizations proposed in \cite{Moiseev2013}. In Section \ref{proposed} we propose novel reduced-rank activity indices, prove that they are unbiased, and discuss their spatial resolution as a function of their rank. Section \ref{ne} provides numerical evaluation of the proposed indices using both simulated and real data.

A short preliminary versions of the paper were presented at conferences \cite{Piotrowski2014a,Piotrowski2014c}.

\section{Preliminaries} \label{preli}
\subsection{Notation}
Assume $\x$ to be a vector of real-valued random variables $x_1,\dots,x_k$, each with a finite variance. The expected value of $\x$ is denoted by $\ev{\x}$ and the covariance matrix of $\x$ is denoted by $\cov{\x}$. $\I_l$ stands for identity matrix of size $l$, while $\I_l^r$ stands for a diagonal matrix of size $l$ in which the first $r$ entries at the main diagonal are equal to 1 and all other entries are equal to~0. By $[\A]_{i,j}$ we denote the $i\times j$ principal submatrix of a matrix $\A.$ We call $\A$ ill-conditioned if some of its singular values are close to zero. By $tr\{\A\}$ we denote the trace of a square matrix $\A.$ Let $\A$ be similar to a symmetric matrix, i.e., $\A=\Pre^{-1}\S\Pre$, where $\Pre$ is invertible and $\S$ symmetric. Then, $\lambda(\A)$ denotes the vector of eigenvalues of $\A$ organized in non-increasing order. We assume that all eigenvalue decompositions considered have eigenvalues organized in non-increasing order. Let $\A$ and $\B$ be symmetric matrices. Then, relation $\A\succeq \B$ denotes Loewner ordering, i.e., $\A-\B$ is positive semidefinite \cite{Horn1985}. Finally, for a given positive (semi)definite matrix $\A$, we denote by $\A^{1/2}$ the unique matrix such that $\A^{1/2}\A^{1/2}=\A$ and $\A^{1/2}$ is also positive (semi)definite \cite{Horn1985}.

The position of a single dipole source is a triple of Cartesian coordinates $\theta$. We denote by $\bth=(\theta_1,\dots,\theta_l)$ an $l$-tuple of Cartesian coordinates of $l$ sources, where $\theta_i$ denotes the position of the $i$-th source, $i=1,\dots,l$.\footnote{See also \textbf{Remark 1} below.} The lead field matrix establishing electromagnetic field propagation model between $l$ dipole sources and $m$ sensors is defined as $\H(\bth)=(h(\theta_1),\dots,h(\theta_l))\in\mathbb{R}^{m \times l}$, where $h(\theta_i)\in\mathbb{R}^{m}$ is the lead field vector of the $i$-th source for $i=1,\dots,l.$ The set of columns of $\H(\bth)$ is denoted by $\{\H(\bth)\}.$ The positions of the $l_0$ sources active during the measurement are denoted by $\bth_0=(\theta_{0,1},\dots,\theta_{0,l_0})$. To simplify notation, explicit dependence on~$\bth$ is omitted whenever possible and values evaluated at~$\bth_0$ are denoted by adding the $0$ subscript, e.g., the lead field matrix evaluated at $\bth_0$ is denoted by $\H_0:=\H(\bth_0)$.

\subsection{EEG/MEG Forward Model} \label{eegmeg}
We consider $l_0$ dipole sources of brain electrical activity and EEG/MEG measurements obtained with $m>l_0$ sensors. We assume that the sources' positions $\bth_0$ are fixed during the measurement period. Then, the $m \times 1$ random vector $\y$ composed of the measurements at a given time instant can be modeled as \cite{VanVeen1997,Mosher1999}:
\begin{equation}
\label{model}
\y=\H_0\q_0+\n,
\end{equation}
where $\H_0:=\H(\bth_0)\in\sgen{m}{l_0}$ is the array response (lead field) matrix corresponding to sources at locations $\bth_0$, $\q_0\in\sgens{l_0}$ is a random vector representing electric/magnetic dipole moments at $\bth_0$, and random vector $\n\in\sgens{m}$ expresses background activity along with noise measured at the sensors.

We assume that the lead field matrix $\H(\bth)$ has full column rank for any set of sources $\bth$ \cite{Sekihara2008}. We also assume that $\q_0$ and $\n$ are zero-mean weakly stationary stochastic processes, covariance matrices $\cov{\q_0}:=\Q$ and $\cov{\n}:=\N$ are positive definite, and that $\q_0$ and $\n$ are uncorrelated. This implies in particular that the covariance matrix of observed signal is also positive definite and is of the form $$\cov{\y}:=\R=\H_0\Q\H_0^t+\N.$$

\begin{remark}
To simplify derivations, in this paper we consider explicitly model (\ref{model}) with constrained orientations of the sources. However, the results of this paper apply equally to the case of unconstrained orientations. In such a case, parametrization of a given source would include not only its position $\theta_i$, but also the unit orientation vector $u_i$. 
\end{remark}  

\subsection{Key Properties of Activity Indices}
\begin{definition} \label{nai}
Define by $\{\H(\bth)\}$ the set of columns (lead fields) of a lead field matrix $\H(\bth)\in\sgen{m}{l}$ of $l$ sources. 
We call a function $f:\{\H(\bth)\}\to\mathbb{R}_+$ an activity index. We say that an activity index $f$ is unbiased if~\cite{Sekihara2008,Moiseev2011}
\begin{equation} \label{unbiasedness}
\{\H(\bth_0)\}\in\bigg\{\arg\max_{\{\H(\bth)\}} f\Big(\{\H(\bth)\}\Big)\bigg\}.
\end{equation}
\end{definition}
A few remarks on Definition \ref{nai} are in place here.
\begin{itemize}
\item An argument of an activity index $f$ is the set of lead fields $\{\H(\bth)\}$ of arbitrary number of sources $l.$ 
\item In the above definition, a lead field $h(\theta)\in\sgens{m}$ serves as a unique identifier of a source at $\theta.$ This is justified by the linear independence of columns of a lead field matrix $\H(\bth)$ for any set of sources $\bth$ \cite{Sekihara2008}. Thus, we may simplify below the notation of an activity index and represent its argument by $\bth$ in place of $\{\H(\bth)\}.$
\item It should be also noted that $\bth_0$ maximizes an unbiased activity index without any additional assumptions about $\Q$ and~$\N$, and in particular, without any assumptions regarding the signal-to-noise ratio.
\item Finally, we note that the set of maximizers of an unbiased activity index may contain elements other than $\bth_0$, as the inverse problem in EEG/MEG is inherently ill-posed~\cite{Helmholtz1853}.
\end{itemize}  

\begin{definition} \label{spatial_res_df}
Consider two unbiased activity indices $f_1$ and $f_2.$ We say that $f_1$ has higher spatial resolution than~$f_2$ if \cite{Sekihara2008,Moiseev2011}
\begin{equation} \label{spatial_res}
\forall\bth\ f_1(\bth)\leq f_2(\bth)\quad\text{and}\quad f_1(\bth_0)=f_2(\bth_0).
\end{equation}
\end{definition}

\subsection{MAI and $MPZ$ Unbiased Activity Indices} \label{MAIMPZ}
Inverse solutions based on spatial filtering seek an estimate of the source vector $\widehat{\q}(\bth)$ in the form $\widehat{\q}=\W(\bth)\y$, where $\W$ is a (l x m) weights matrix. Specifically, for the MCMV filter, which is a multi-source version of a traditional LCMV beamformer \cite{Frost1972,VanVeen1997,Sekihara2008}, the weights are given by the formula (see \ref{opt}) \begin{equation} \label{lcmv}
\W(\bth)=\S(\bth)^{-1}\H(\bth)^t\R^{-1},
\end{equation}
where $\S(\bth)\in\sgen{l}{l}$ is a positive definite matrix of the form
\begin{equation} \label{S}
\S:=\S(\bth)=\H(\bth)^t\R^{-1}\H(\bth):=\H^t\R^{-1}\H\succ 0.
\end{equation} 
The work \cite{Moiseev2011} introduced, among others, two unbiased activity indices for the above filter, which extend previously known single source indices to the multi-source case. Physically those indices represent certain signal to noise ratios for the reconstructed source activity estimated by the filter. 

The signal-to-noise ratios in question are obtained as follows: the covariance matrix of the signal $\widehat{\q}=\W(\bth)\y$ reconstructed using MCMV filter is
\begin{equation} \label{Wy}
\cov{\W(\bth)\y}=\W(\bth)\R\W(\bth)^t=\S^{-1}.
\end{equation}
Assume now that only background activity along with noise (i.e., $\n$ in (\ref{model})) is observed. Then, $\R$ reduces to $\N$, and the MCMV filter becomes
\begin{equation}
\W_N(\bth)=\G(\bth)^{-1}\H(\bth)^t\N^{-1},
\end{equation}
where $\G(\bth)\in\sgen{l}{l}$ is a positive definite matrix of the form
\begin{equation} \label{G}
\G:=\G(\bth)=\H(\bth)^t\N^{-1}\H(\bth):=\H^t\N^{-1}\H\succ 0.
\end{equation}
Then, one has in such a case that 
\begin{equation} \label{WNn}
\cov{\W_N(\bth)\n}=\W_N(\bth)\N\W_N(\bth)^t=\G^{-1}.
\end{equation}
For a single source case of $l=1$, $\S$ and $\G$ become scalar values and the ratio
\begin{equation} \label{GS}
\cov{\W(\bth)\y}/\cov{\W_N(\bth)\n}-1=\S^{-1}/\G^{-1}-1=\G\S^{-1}-1
\end{equation}
has been known in literature as \emph{neural activity index} since its introduction in the seminal paper \cite{VanVeen1997}.\footnote{In \cite{VanVeen1997}, \emph{neural activity index} was defined without subtracting one from the power ratio $\G\S^{-1}.$ We have done so here to emphasize the relation between single-source indices and their multi-source generalizations introduced below.}

Moreover, assuming full signal $\y$ in model (\ref{model}) is observed, the covariance matrix of the noise projected by the MCMV filter is of the form:
\begin{equation} \label{Wn}
\cov{\W(\bth)\n}=\S^{-1}\H(\bth)^t\R^{-1}\N\R^{-1}\H(\bth)\S^{-1}=\S^{-1}\T\S^{-1},
\end{equation}
where $\T\in\sgen{l}{l}$ is a positive definite matrix such that
\begin{equation} \label{T}
\T:=\T(\bth)=\H(\bth)^t\R^{-1}\N\R^{-1}\H(\bth):=\H^t\R^{-1}\N\R^{-1}\H\succ 0.
\end{equation}  
Then, for a single source case of $l=1$, the ratio
\begin{equation} \label{ST}
  \cov{\W(\bth)\y}/\cov{\W(\bth)\n}-1=
  \S^{-1}/\big(\S^{-1}\T\S^{-1}\big)-1=\S\T^{-1}-1,
\end{equation}
has been known in literature as \emph{pseudo-Z index} \cite{Robinson1999}. 

The work \cite{Moiseev2011} extended single-source activity indices in (\ref{GS}) and (\ref{ST}) by introducing, respectively:
\begin{itemize}
\item \emph{Multi-source Activity Index} ($MAI$) defined in \cite{Moiseev2011} as
\begin{equation} \label{MAI}
MAI(\bth)=tr\{\G\S^{-1}\}-l,
\end{equation}
\item \emph{Multi-source Pseudo-Z index} ($MPZ$) defined in \cite{Moiseev2011} as
\begin{equation} \label{MPZ}
MPZ(\bth)=tr\{\S\T^{-1}\}-l.
\end{equation}
\end{itemize}

It has been proved in \cite{Moiseev2011} that $MAI$ and $MPZ$ activity indices are unbiased in the sense of Definition \ref{nai}. Moreover, it has been also proved that these localizers satisfy the \emph{saturation} property: an increase of the number of sources $l$ above $l_0$ results in the same maximizing values of $MAI$ and $MPZ$ activity indices as those obtained for $l_0.$ For $l$ such that $1\leq l\leq l_0$ the maximizing values of $MAI$ and $MPZ$ activity indices are monotonically increasing with $l$ until $l_0$ is reached. Further on, for the sake of theoretical derivations below, we assume that $l$ is such that $1\leq l\leq l_0$ and that the true number of sources $l_0$ is known.

\begin{remark} \label{saturation}
While the saturation property suggests a way to determine the actual number of active sources $l_0$ from the data, in practice this may be not the best approach; please refer to  \cite{Moiseev2011} for a detailed discussion of this issue.  
\end{remark}  

It has been also established in \cite{Moiseev2011} that $MPZ$ has higher spatial resolution than $MAI.$ See \ref{mai_mpz_res_pd} for a sketch of proof of this fact.

\subsection{Extensions of $MAI$ and $MPZ$ Activity Indices} \label{ext}
An interesting extension of results in \cite{Moiseev2011} is provided in~\cite{Moiseev2013}, where it has been proved that all combinations of eigenvalues of $\G\S^{-1}$ and all combinations of eigenvalues of $\S\T^{-1}$ have stationary points at $\bth_0$, extending in this sense $MAI$ and $MPZ$ localizers. Moreover, Monte-Carlo simulations presented in \cite{Moiseev2013} showed that the sum of the largest eigenvalues yielded the most stable results.

Let $r\in\mathbb{N}$ be a natural number less than $l_0.$ Following~\cite{Moiseev2013}, we consider the extensions of $MAI$ and $MPZ$ activity indices defined~as
\begin{equation} \label{MAI_ext}
  MAI_{ext}(\bth,r)=\sum_{i=1}^{r}\lambda_i(\G\S^{-1})-r,\ 1\leq r\leq l_0,
\end{equation}
and
\begin{equation} \label{MPZ_ext}
  MPZ_{ext}(\bth,r)=\sum_{i=1}^{r}\lambda_i(\S\T^{-1})-r,\ 1\leq r\leq l_0,
\end{equation}
respectively.

We note that $MAI_{ext}$ and $MPZ_{ext}$ activity indices are not unbiased in the sense of Definition \ref{nai}. However, due to the fact that all combinations of eigenvalues of $\G\S^{-1}$ and all combinations of eigenvalues of $\S\T^{-1}$ have stationary points at $\bth_0$, the $MAI_{ext}$ and $MPZ_{ext}$ activity indices are called \emph{unbiased in a broad sense} in~\cite{Moiseev2013}. 

\section{Proposed Activity Indices} \label{proposed}
A motivation for introducing multi-source $MAI$ and $MPZ$ localizers defined in Section \ref{MAIMPZ} is that performance of the traditional single source (LCMV) based ones is known to be poor if sources are strongly correlated, or the lead field matrix is ill-conditioned as a result of source proximity, or when strong background activity and measurement noise is present - see, e.g., \cite{VanVeen1997, Piotrowski2014b, Piotrowski2018s} and references therein. It is therefore intriguing to investigate how such conditions affect spatial resolution of the $MAI$ and $MPZ$ activity indices.

To this end, we shall consider below the following forward model equivalent to (\ref{model}):
\begin{equation} \label{model_eq}
\y=\H_0\Q^{1/2}\Q^{-1/2}\q_0+\n=\H_0'\q'_0+\n,
\end{equation}  
with $\H_0'=\H_0\Q^{1/2}$ and $\q'_0=\Q^{-1/2}\q_0$ such that $\cov{\q'_0}=\I_{l_0}.$ Then, letting $\H':=\H([\Q]_{l\times l})^{1/2}$ we may define $\S'$, $\G'$, and $\T'$ as above, by using $\H'$ in place of $\H$, e.g., $\S':=(\H')^t\R^{-1}\H'.$

\subsection{Reduced-Rank Approach}
One of the known ways in signal processing to remedy the drawbacks of the LCMV/MCMV-type filters is to apply reduced-rank solutions \cite{Scharf1991, Stoica1996, Werner2006}. Following this line, in this paper we develop activity indices which are reduced-rank extensions of $MAI$ and $MPZ$ localizers.

We note first that the MCMV filter designed for model (\ref{model_eq}) is of the following form:
\begin{equation} \label{lcmv'}
\W'(\bth):=\S'(\bth)^{-1}\H'(\bth)^t\R^{-1}=\Q^{-1/2}\W(\bth).
\end{equation}
Let $r\in\mathbb{N}$ be such that $1\leq r\leq l_0.$ Below, we propose activity indices based on the MV-PURE filter \cite{Piotrowski2008}, given in terms of model (\ref{model_eq}) as
\begin{equation} \label{mvp}
\W'_{RR}(\bth)=\Pre^{(r)}_{\S'}\W'(\bth),
\end{equation}  
where by $\Pre^{(r)}_{\S'}=\V\I_l^r\V^t$ we denote orthogonal projection matrix onto subspace spanned by eigenvectors corresponding to the $r$ largest eigenvalues of~$\S'$ given with eigenvalue decomposition $\S'=\V\La \V^t.$ The MV-PURE filter is a natural reduced-rank extension of the MCMV filter, \emph{cf.} optimization problems given in \ref{opt}, for which $\W'(\bth)$ and $\W'_{RR}(\bth)$ are the solutions.

Consider now the signal-to-noise ratios of the MV-PURE filter. Owing to the symmetry of $\Pre^{(r)}_{\S'}$, the expressions corresponding to (\ref{Wy}) and (\ref{WNn}) are, respectively,
\begin{equation}
\cov{\W'_{RR}(\bth)\y}=\Pre^{(r)}_{\S'}\W'(\bth)\R(\Pre^{(r)}_{\S'}\W'(\bth))^t=\Pre^{(r)}_{\S'}(\S')^{-1}\Pre^{(r)}_{\S'},
\end{equation}
and
\begin{equation}
\cov{\W'_{{RR}_N}(\bth)\n}=\Pre^{(r)}_{\S'}\W'_N(\bth)\N(\Pre^{(r)}_{\S'}\W'_N(\bth))^t=\Pre^{(r)}_{\S'}(\G')^{-1}\Pre^{(r)}_{\S'},
\end{equation}  
where $\W'_{{RR}_N}:=\Pre^{(r)}_{\S'}\W'_N$ with
\begin{equation*}
\W'_N(\bth):=\G'(\bth)\H'(\bth)^t\N^{-1}=\Q^{-1/2}\W_N(\bth),
\end{equation*}
\emph{cf.} (\ref{lcmv'}).\footnote{If only background activity $\n$ is observed, we could also use the alternative form of the filter $\W'_{{RR}_N}:=\Pre^{(r)}_{\G'}\W'_N.$ However, this would lead to a more involved form of the proposed index, with both $\Pre^{(r)}_{\G'}$ and $\Pre^{(r)}_{\S'}$ present.} Then, in view of idempotency of $\Pre^{(r)}_{\S'}$, and the fact that 
\begin{multline} \label{inout}
\Pre^{(r)}_{\S'}(\S')^{-1}\Pre^{(r)}_{\S'}=\V\I_l^r\V^t\V\La^{-1} \V^t\V\I_l^r\V^t=\\
\V\I_l^r\La^{-1}\I_l^r\V^t=\V\La^{-1}\I_l^r\V^t=\V\La^{-1}\V^t\V\I_l^r\V^t=(\S')^{-1}\Pre^{(r)}_{\S'},
\end{multline}
and proceeding analogously as in \cite{Moiseev2011} by generalizing the ratio $$\cov{\W'_{RR}(\bth)\y}/\cov{\W'_{{RR}_N}(\bth)\n}$$ to the multi-source case (\emph{cf.} (\ref{GS})),
we obtain the following reduced-rank extension of the $MAI$ activity index (\ref{MAI}):
\begin{multline} \label{RRMAI}
MAI_{RR}(\bth,r)=tr\{\Pre^{(r)}_{\S'}\G'\Pre^{(r)}_{\S'}\Pre^{(r)}_{\S'}(\S')^{-1}\Pre^{(r)}_{\S'}\}-r=\\
tr\{\G'\Pre^{(r)}_{\S'}(\S')^{-1}\Pre^{(r)}_{\S'}\}-r:=tr\{\G'(\S')^{-1}\Pre^{(r)}_{\S'}\}-r,\ 1\leq r\leq l_0.
\end{multline}
Similarly, for the generalization of the $MPZ$ activity index to the reduced-rank case, we consider now the expression of the covariance matrix of the noise projected by the MV-PURE filter, \emph{cf.} (\ref{Wn}):
\begin{equation} 
\cov{\W'_{RR}(\bth)\n}=\Pre^{(r)}_{\S'}(\S')^{-1}\H'(\bth)^t\R^{-1}\N\R^{-1}\H'(\bth)(\S')^{-1}\Pre^{(r)}_{\S'}.
\end{equation}  
Then, from (\ref{ST}), and in view of the above derivation for $MAI_{RR}$, it is clear that the generalized ratio $\cov{\W'_{RR}(\bth)\y}/\cov{\W'_{RR}(\bth)\n}$ defining $MPZ_{RR}$ activity index will be of the form
\begin{equation} \label{RRMPZ}
MPZ_{RR}(\bth,r):=tr\{\S'(\T')^{-1}\Pre^{(r)}_{\S'}\}-r,\ 1\leq r\leq l_0.
\end{equation}

\subsection{Key Properties of $MAI_{RR}$ and $MPZ_{RR}$ Activity Indices} \label{properties}
We begin this section with the following lemma.
\begin{lemma} \label{L1}
Let $r$ be a natural number less than $l_0.$ Then
\begin{equation} \label{l1}
MAI_{RR}(\bth,r)\leq\sum\limits_{i=1}^{r}\lambda_{i}(\R\N^{-1})-r=\sum\limits_{i=1}^{r}\lambda_{i}(\G'_{0}).
\end{equation}  
\end{lemma}  
\pd See \ref{L1_pd}.

We are now ready to prove the following theorem.
\begin{theorem} \label{Th1}
Let $r$ be a natural number less than $l_0.$ Then, $MAI_{RR}$ and $MPZ_{RR}$ activity indices are unbiased. Moreover, $MPZ_{RR}$ has higher spatial resolution than $MAI_{RR}$ for the same rank constraint $r.$ In particular,  one has
\begin{equation} \label{th1}
MAI_{RR}(\bth_0,r)=MPZ_{RR}(\bth_0,r)=\sum_{i=1}^{r}\lambda_{i}(\G'_{0}).
\end{equation}
\end{theorem}
\pd See \ref{Th1_pd}.

Theorem \ref{Th1} introduces two families of unbiased activity indices parameterized by rank. The important questions are: when it is useful to lower the rank of a given unbiased activity index, and how this parameter should be selected in practice. To provide insight into these questions, we begin with the following propositions.

\begin{proposition} \label{trailing}
Let $r$ be a natural number less than $l_0.$ Then
\begin{equation} \label{epsilon}
MAI(\bth_0)-MAI_{RR}(\bth_{0},r)=MPZ(\bth_0)-MPZ_{RR}(\bth_{0},r)=\sum_{i=r+1}^{l_0}\lambda_{i}(\G'_{0}).
\end{equation}
\end{proposition}
\pd The proof is obtained immediately from (\ref{th1}) in Theorem \ref{Th1}. 

\begin{proposition} \label{ineqs}
Let $r$ be a natural number less than $l_0.$ Then
\begin{equation} \label{ineqsMAI}
MAI_{RR}(\bth,r)\leq MAI_{RR}(\bth,r+1)\leq MAI_{RR}(\bth;l_0)=MAI(\bth),
\end{equation}
and similarly
\begin{equation} \label{ineqsMPZ}
MPZ_{RR}(\bth,r)\leq MPZ_{RR}(\bth,r+1)\leq MPZ_{RR}(\bth;l_0)=MPZ(\bth).
\end{equation}
\end{proposition}
\pd See \ref{ineqs_pd}.

\begin{figure}[h] 
\centering
\includegraphics[width=0.6\linewidth]{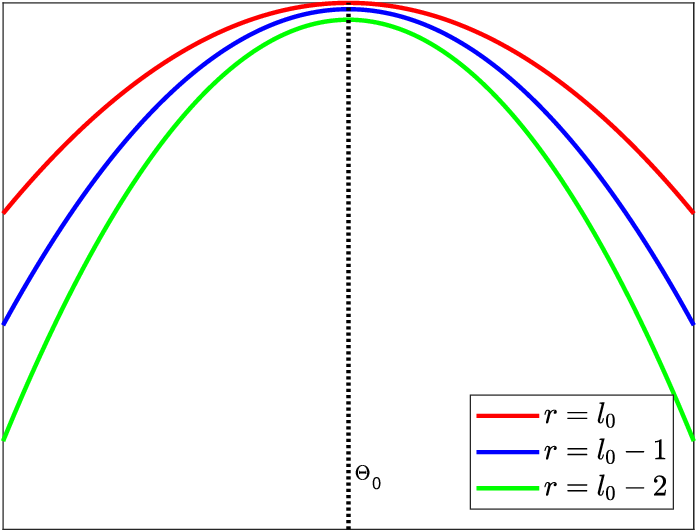}
\caption{Theoretical graphs of $MAI_{RR}(\bth,r)$ and $MPZ_{RR}(\bth,r)$ for $r\in\{l_0-2,l_0-1,l_0\}.$ In the situation depicted in this figure, the trailing eigenvalues $\lambda_{l_0-1}(\G'_{0})$ and $\lambda_{l_0}(\G'_{0})$ of $\G'_{0}$ are close to zero, resulting in approximately the same maximum value at $\bth_0$ of $MAI_{RR}(\bth,r)$ and $MPZ_{RR}(\bth,r)$ for $r\in\{l_0-2,l_0-1,l_0\}$, \emph{cf.} (\ref{epsilon}).} \label{spatial_res_fig}
\end{figure}

From (\ref{epsilon}), it is seen in particular that if the trailing eigenvalues $$\lambda_{r+1}(\G'_{0}),\dots,\lambda_{l_0}(\G'_{0})$$ of $\G'_{0}$ are close to zero for some $r<l_0$, the maximizing value will be almost identical for activity indices of ranks $r,\dots,l_0.$ Then, taking into account Definition \ref{spatial_res_df}, we conclude from (\ref{ineqsMAI}) and (\ref{ineqsMPZ}) that in such a case, the activity index of rank $r$ provides approximately the highest spatial resolution among activity indices of ranks $r,\dots,l_0.$ Fig.~\ref{spatial_res_fig} illustrates such situation.

In view of the above analysis, we should find now when the trailing eigenvalues $\lambda_{r+1}(\G'_{0}),\dots,\lambda_{l_0}(\G'_{0})$ of $\G'_{0}$ are close to zero for some $r<l_0.$ To this end, we note first that
\begin{multline} \label{MAImotive}
\lambda_i(\G'_0)=\lambda_i(\Q^{1/2}\H_0^t\N^{-1}\H_0\Q^{1/2})=\\\lambda_i(\Q\H_0^t\N^{-1}\H_0)=\lambda_i(\H_0^t\N^{-1}\H_0\Q),\ 1\leq i\leq l_0.
\end{multline}
Then, from Fact \ref{Lidskiis} in \ref{kru} we obtain, upon setting $k=1$ therein, that for $i\in\{1,\dots,l_0\}$ one has
\begin{equation}
\lambda_i(\G'_0)\leq\lambda_i(\Q)\lambda_1(\H_0^t\N^{-1}\H_0),
\end{equation}
and similarly
\begin{equation}
\lambda_i(\G'_0)\leq\lambda_1(\Q)\lambda_i(\H_0^t\N^{-1}\H_0).
\end{equation}
It is seen, therefore, that $\sum_{i=r+1}^{l_0}\lambda_{i}(\G'_{0})$ in (\ref{epsilon}) will be close to $0$ for some $r<l_0$, provided $\lambda_{r+1}(\Q),\dots,\lambda_{l_0}(\Q)$ or $\lambda_{r+1}(\H_0^t\N^{-1}\H_0),\dots,\lambda_{l_0}(\H_0^t\N^{-1}\H_0)$ are close to zero. The former occurs if the activity is heavily correlated among some sources, and the latter if the lead field matrix $\H_0$ is ill-conditioned, and especially if the eigenvalues of $\N$ are large, implying low signal-to-noise ratio. From our experience with EEG/MEG forward models, $\H_0$ is likely to be ill-conditioned if some of the active sources are closely positioned. 

The remaining question is how the eigenvalues of $\G'_0$ in (\ref{MAImotive}) can be estimated in practice, as neither $\Q$ nor $\H_0$ are observable. However, we note that
\begin{multline} \label{note}
\lambda_i(\R\N^{-1})=\lambda_i\big((\H_0\Q\H_0^t+\N)\N^{-1}\big)=\\\lambda_i(\H_0\Q\H_0^t\N^{-1}+\I_m)=\lambda_i(\Q\H_0^t\N^{-1}\H_0)+1=\\
\lambda_i(\G'_0)+1,\ 1\leq i\leq l_0.
\end{multline}  
The covariance matrices $\R$ and $\N$ can be estimated from the available measurements, with the latter obtainable from pre-stimulus interval of the experiment or directly from the estimated spectrum of $\R$ if white noise is assumed \cite{Sekihara2008}. 

We close this section with a proposition showing that the proposed activity indices would have higher spatial resolution, for the same rank constraint $r$, than the extensions of MAI and MPZ activity indices of the forms (\ref{MAI_ext}) and (\ref{MPZ_ext}), even if the latter were unbiased.

\begin{proposition} \label{superiority_ext}
Let $r$ be a natural number less than $l_0$, and consider $MAI_{ext}$ and $MPZ_{ext}$ localizers of the forms (\ref{MAI_ext}) and (\ref{MPZ_ext}), respectively. Then
\begin{equation} \label{Ext1}
MAI_{RR}(\bth,r)\leq MAI_{ext}(\bth,r),
\end{equation}
and
\begin{equation} \label{Ext2}
MPZ_{RR}(\bth,r)\leq MPZ_{ext}(\bth,r).
\end{equation}
Moreover,
\begin{multline} \label{Ext3}
MAI_{RR}(\bth_0,r)=MPZ_{RR}(\bth_0,r)=\\ MAI_{ext}(\bth_0,r)=MPZ_{ext}(\bth_0,r)=\sum_{i=1}^{r}\lambda_{i}(\G'_{0}).
\end{multline}
\end{proposition}
\pd See \ref{superiority_ext_pd}.

\section{Iterative Implementation and Numerical Examples} \label{ne}
To test the properties of the activity indices introduced in this paper for the EEG/MEG forward model introduced in Section \ref{eegmeg}, we present below their sample applications to the EEG/MEG source localization problem on a simulated EEG and real MEG data. In both cases, we use the following computationally efficient iterative implementation of the proposed source localization methods.

\subsection{Iterative Implementation} \label{iterative}
In view of our original assumption that the number of active sources $l_0$ is known, it may seem at first glance that it is sufficient to evaluate multi-source indices for $l=l_0$ only. However, in practice, $l_0$ is usually estimated from the data, or using some stopping criterion for the applied iterative algorithm. This determines an effective number of sources $\tilde{l}_0$ which is an approximation of the true~$l_0.$ Moreover, regardless whether $l_0$ or $\tilde{l}_0$ is available, a brute-force evaluation of a multi-source index over multidimensional domain is in most cases infeasible, as it contains ${s}\choose{l_0}$ \Big(or ${s}\choose{\tilde{l}_0}$\Big) elements, where $s$ is the total number of all possible locations of active sources.

Therefore, for the numerical implementation of the proposed indices, we follow \cite{Moiseev2011} and adapt the iterative procedure used previously for MAI and MPZ activity indices.\footnote{The iterative procedure used in \cite{Moiseev2011} considers also finding orientation of the dipole sources. We have omitted this step in the proposed implementation assuming orientations known from the forward model.} This procedure discovers locations of sources sequentially, beginning with the strongest source. As described in \cite{Moiseev2011}, such approach meets the specifics of the inverse problem in EEG/MEG well, and is capable of closely approximating the optimal solution. For the sake of notational compliance of the presentation, we have used $l_0$ in the following pseudocode, as its logic is unchanged if $l_0$ is replaced by $\tilde{l}_0.$

\begin{algorithm}[h] 
  \caption{Iterative discovery of sources using multi-source activity index}
  \label{A1}
  \begin{algorithmic}[1]
    \renewcommand{\algorithmicrequire}{\textbf{Input: parameters $l_0$, $s$, $r$, activity index $f(\bth,r)$}}
    \REQUIRE in
    % \ENSURE  out
    \\ \textit{Initialisation}
    \STATE $\bth_0=\{\varnothing\}$
    \\ \textit{LOOP Process}
    \FOR {$l=1$ to $l_0$}
    \FOR {$i=1$ to $s$}
    \STATE set $\bth_i=\bth_0\cup\{\theta_i\}$
    \ENDFOR
    \IF {$l\leq r$}
    \STATE $\bth_0=\underset{\bth_1,\dots,\bth_s}{\arg\max}f(\bth_i,l)$ 
    \ELSE
    \STATE $\bth_0=\underset{\bth_1,\dots,\bth_s}{\arg\max}f(\bth_i,r)$ 
    \ENDIF
    \ENDFOR
    \RETURN $\bth_0$
  \end{algorithmic}
\end{algorithm}

We observe that the iterative process described in Algorithm~\ref{A1} produces for $l\leq r$ exactly the same output as its version used in \cite{Moiseev2011}, since for these iterations the activity indices of (full) rank $l$ are used. However, for $l>r$, the reduced-rank approach is used in the above implementation, and thus is aimed at finding weaker sources at later iterations. In such a case, one needs in principle to estimate $[\Q]_{l\times l}$ for $\H'$ needed for $MAI_{RR}$ and $MPZ_{RR}$ localizers, \emph{cf.} (\ref{model_eq}), (\ref{RRMAI}) and (\ref{RRMPZ}). We will present an estimation method for $[\Q]_{l\times l}$ in Section \ref{assr} and use it on real data, where the weaker sources may exhibit correlated activity. For the sake of simulations, however, we present a simplified method which assumes that for iterations $l>r$ the activity of sources is uncorrelated with other sources, which frees us from the need of computing the estimate of $[\Q]_{l\times l}$ at these iterations, thus reducing the computational effort.

We note that Algorithm \ref{A1} requires at each iteration only $s$ evaluations of a given multi-source activity index. The computational complexity of full-rank and reduced-rank indices is of the same order at each iteration. It shall be also emphasized that the above algorithm is just an example of an iterative scheme for computation of multi-source indices. Indeed, another iterative scheme with slightly more computational steps needed at each iteration has been proposed in~\cite{Herdman2018}. 

\subsection{EEG Numerical Simulations} \label{nene}
The simulations presented in this section use open-source EEG/MEG spatial filtering supFunSim toolbox available for download at
\newline \url{https://github.com/nikadon/supFunSim} \cite{Rykaczewski2020} to generate simulated EEG source signals and calculate EEG lead fields in order to obtain realistic simulated EEG electrode signals. This toolbox has been also used to generate numerical simulations in \cite[Section VI]{Piotrowski2018s} and the EEG signal in \cite{Kono2019}. We describe below the key properties of this framework relevant to simulations considered in this paper. 

Generation of time series in source space for bioelectrical activity of brains'
cortical and subcortical regions is conducted using separate
MVAR models (of order 6) for the active sources $\q_0$ and the biological background noise \cite{Rykaczewski2020}, see also \cite{Korzeniewska2003} for a similar approach.
We add Gaussian uncorrelated noise to the time series in sensor space to model intrinsic sensor noise. Moreover, the coefficient matrix of MVAR model used to generate $\q_0$ was multiplied by a non-diagonal mask matrix yielding correlated activity among active sources.

For solution of the forward problem, supFunSim uses the FieldTrip toolbox \cite{FieldTrip2011} for generation of 
the volume conduction model (VCM) needed to obtain EEG lead fields. For the results presented herein, we arbitrarily select:
\begin{itemize}
\item \emph{HydroCel Geodesic Sensor Net} utilizing 128 channels as EEG cap layout.
\item Regions of interest (ROIs) corresponding to specific cortex patches, and whose geometry is reconstructed and parcellated from the detailed cortical surface by means of the Brainstorm toolbox \cite{Brainstorm2011}. We selected ROIs by
their anatomical description in Destrieux and Desikan-Killiany atlases \cite{FreeSurferCortex2004,FreeSurferCortex2006}. Surface parcellation was prepared using freely available
FreeSurfer Software
Suite \cite{FreeSurfer1999} such that each ROI is comprised of a triangular mesh and fully characterized by its nodes, from which candidates for source position and orientation (orthogonal to the mesh) will be randomly drawn.
\item Both thalami modeled jointly as a single triangulated mesh containing node candidates
for a random selection of subcortical bioelectrical activity. Also here, the orientation of sources is chosen as orthogonal to the mesh surface.
\end{itemize}

The above procedure resulted in $s=16144$ candidate positions. We considered 9 active sources, such that 3 of them were in fixed and closely positioned locations, with the locations of the remaining 6 randomly assigned among ROIs nodes, ensuring even distribution of sources among ROIs. The signal-to-noise ratio (SNR) was defined as $SNR=||\H_0\q_0||/||\n||.$ For each of the SNR ratio considered, we conducted 10 simulation runs. In each run, locations of 6 out of 9 active sources were randomly chosen in accordance to the above mentioned scheme, and a new MVAR model was used to generate the source signals.

Each simulation run contained a realization of the MVAR process with 1000 samples, where:
\begin{itemize}
\item The first half (i.e., the first 500 samples) of each trial is interpreted as pre-task/stimulus
activity and represents noise signal $\n.$ The estimate of noise covariance matrix $\bm{N}$ is obtained from this section of the signals as a finite sample estimate.
\item The second half of each trial is comprised of both noise and activity of sources $\q_0.$ The estimate of signal covariance matrix $\bm{R}$ is obtained from this second section of the signals as a finite sample estimate. 
\end{itemize}

We selected for evaluation the following activity indices, all of which are based on the activity indices introduced in Sections \ref{preli} and \ref{proposed}, adjusted for use in the iterative computation scheme considered in this section:
\begin{itemize}
\item \texttt{MAI} activity index defined at $l$-th iteration as:
  \begin{equation} \label{MAI_it}
    \texttt{MAI}=tr\{\G\S^{-1}\}-l,\ 1\leq l\leq l_0.
  \end{equation}
\item \texttt{MAI}$_{ext}$ activity index defined at $l$-th iteration as:
  \setstretch{2}
  \begin{equation}
    \texttt{MAI}_{ext}=\left\{
    \begin{array}{l}
      \sum_{i=1}^{l}\lambda_i(\G\S^{-1})-l,\ 1\leq l\leq r,\\
      \sum_{i=1}^{r}\lambda_i(\G\S^{-1})-r,\ r<l\leq l_0.\\
    \end{array}\right.
  \end{equation}
  \setstretch{1}
  We note that \texttt{MAI}$_{ext}$=\texttt{MAI} for $1\leq l\leq r.$
\item \texttt{MAI}$_{RR-I}$ activity index defined at $l$-th iteration as:
  \setstretch{2}
  \begin{equation} \label{MAI_RR_I_it}
    \texttt{MAI}_{RR-I}=\left\{
    \begin{array}{l}
      tr\{\G\S^{-1}\}-l,\ 1\leq l\leq r,\\
      tr\{\G\S^{-1}\Pre^{(r)}_{\S}\}-r,\ r<l\leq l_0.\\ 
    \end{array}\right.
  \end{equation}
  \setstretch{1}
  We note that \texttt{MAI}$_{RR-I}$=\texttt{MAI} for $1\leq l\leq r.$
\item \texttt{MPZ} activity index defined at $l$-th iteration as:
  \begin{equation} 
    \texttt{MPZ}=tr\{\S\T^{-1}\}-l,\ 1\leq l\leq l_0.
  \end{equation}
\item \texttt{MPZ}$_{ext}$ activity index defined at $l$-th iteration as:
  \setstretch{2}
  \begin{equation}
    \texttt{MPZ}_{ext}=\left\{
    \begin{array}{l}
      \sum_{i=1}^{l}\lambda_i(\S\T^{-1})-l,\ 1\leq l\leq r,\\
      \sum_{i=1}^{r}\lambda_i(\S\T^{-1})-r,\ r<l\leq l_0.\\
    \end{array}\right.
  \end{equation}
  \setstretch{1}
  We note that \texttt{MPZ}$_{ext}$=\texttt{MPZ} for $1\leq l\leq r.$
\item \texttt{MPZ}$_{RR-I}$ activity index defined at $l$-th iteration as:
  \setstretch{2}
  \begin{equation} \label{MPZ_RR_I_it}
    \texttt{MPZ}_{RR-I}=\left\{
    \begin{array}{l}
      tr\{\S\T^{-1}\}-l,\ 1\leq l\leq r,\\
      tr\{\S\T^{-1}\Pre^{(r)}_{\S}\}-r,\ r<l\leq l_0.\\ 
    \end{array}\right.
  \end{equation}
  \setstretch{1}
  We note that \texttt{MPZ}$_{RR-I}$=\texttt{MPZ} for $1\leq l\leq r.$
\end{itemize}

In the above formulation, only \texttt{MAI} and \texttt{MPZ} activity indices are unbiased in the sense of Definition \ref{nai}. The reduced-rank activity indices \texttt{MAI}$_{ext}$ and \texttt{MPZ}$_{ext}$ are unbiased in broad sense, while \texttt{MAI}$_{RR-I}$ and \texttt{MPZ}$_{RR-I}$ are unbiased under simplifying assumption of uncorrelated sources. Then, the analysis of key properties of $MAI_{RR}$ and $MPZ_{RR}$ activity indices made in Section \ref{properties} may be adjusted to current iterative scheme under this assumption. In particular, we selected their rank $r$ used in Algorithm \ref{A1} as follows:\footnote{We used the same rank for \texttt{MAI}$_{ext}$ and \texttt{MPZ}$_{ext}$ activity indices as well.} first, we estimated the eigenvalues of $\lambda_i(\G'_0)$ for $1\leq i\leq l_0$ from the $l_0$ largest eigenvalues of $\R\N^{-1}$, \emph{cf.} (\ref{note}). We then considered normalized eigenvalues as $\lambda'_i(\R\N^{-1}):=\lambda_i(\R\N^{-1})/\sum_{i=1}^{l_0}\lambda_i(\R\N^{-1})$ for $1\leq i\leq l_0$, and selected $r$ such that
\begin{equation} \label{r}
r=\arg\min_{1\leq l\leq l_0}\sum_{i=1}^l\lambda'_i(\R\N^{-1})>\delta\in(0,1].
\end{equation}
The closer the $\delta$ is to $1$, the less likely it becomes to discard trailing eigenvalues of $\R\N^{-1}$ (and hence, of $\lambda_i(\G'_0)$), thus reducing the possibility of obtaining reduced-rank activity index with increased spatial resolution compared with full-rank activity index, as discussed below Proposition \ref{ineqs} in Section \ref{proposed}. On the other hand, if one chooses too small value of $\delta$, from the same discussion it is seen that in such a case there is no guarantee that the resulting activity index will have better spatial resolution compared with full-rank activity index. In the current simulations, we set $\delta=0.8$ to accommodate the above considerations.

\begin{figure}[p!] 
\centering
\includegraphics[width=0.6\linewidth]{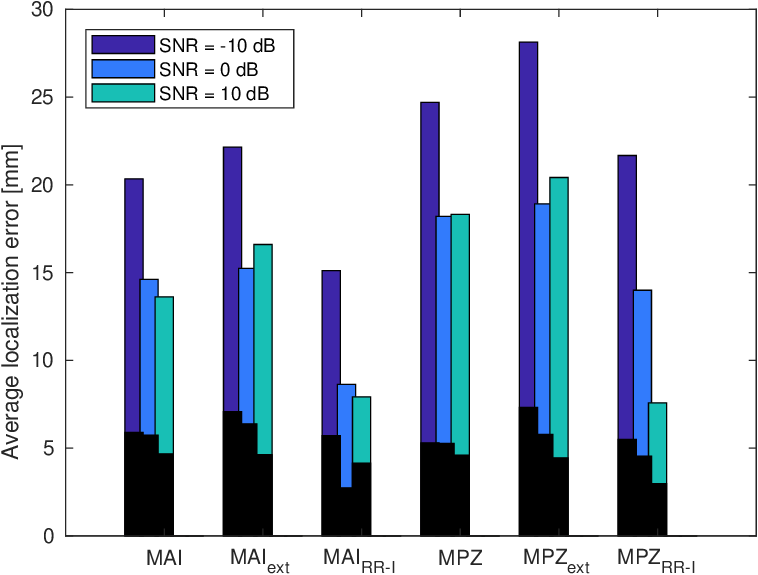}
\caption{Localization error averaged over simulation runs and all iterations of Algorithm~\ref{A1}. Black bars indicate standard deviations of corresponding localization errors.} \label{sim1}
\end{figure}

\begin{table}[p!] 
\small
\centering
\captionsetup{font=small}
\caption{Corresponding $p$-value of right-tailed Mann-Whitney-Wilcoxon test}
\begin{tabular}{r|ll}
\multirow{2}{*}{SNR [dB]} & \multicolumn{2}{c}{hypothesis tested}                             \\ \cline{2-3}
                     & \multicolumn{1}{c}{\texttt{MAI}$>$\texttt{MAI}$_{RR-I}$} & \multicolumn{1}{c}{\texttt{MAI}$_{ext}>$\texttt{MAI}$_{RR-I}$} \\\specialrule{1.5pt}{0pt}{0pt}
-10 & 0.0188  & 0.0086 \\
0 & 0.0046 & 0.0057 \\
10 & 0.007 & 0.0007 \\
\specialrule{1.5pt}{0pt}{0pt}
\multirow{2}{*}{SNR [dB]} & \multicolumn{2}{c}{hypothesis tested}                             \\ \cline{2-3}
                     & \multicolumn{1}{c}{\texttt{MPZ}$>$\texttt{MPZ}$_{RR-I}$} & \multicolumn{1}{c}{\texttt{MPZ}$_{ext}>$\texttt{MPZ}$_{RR-I}$} \\\specialrule{1.5pt}{0pt}{0pt}
-10 & 0.1061 & 0.032 \\
0 & 0.0606 & 0.0378 \\
10 & 0.0002 & 0.0002 \\
\end{tabular}
\small
\end{table}

\begin{figure}[p!] 
\centering
\includegraphics[width=0.6\linewidth]{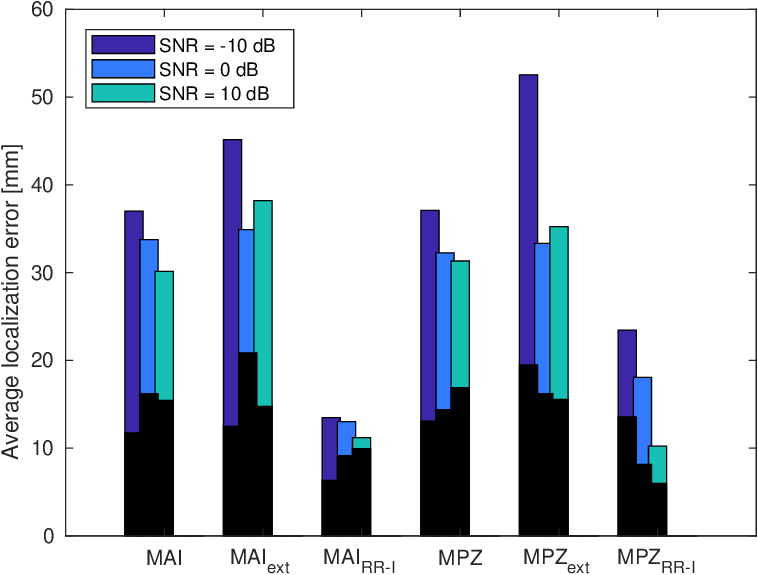}
\caption{Localization error averaged over simulation runs and sources found at last 2 iterations ($l\in\{l_0-1,l_0\}$) of Algorithm~\ref{A1}. Black bars indicate standard deviations of corresponding localization errors.} \label{sim2}
\end{figure}

\begin{table}[p!] 
\small
\centering
\captionsetup{font=small}
\caption{Corresponding $p$-value of right-tailed Mann-Whitney-Wilcoxon test}
\begin{tabular}{r|ll}
\multirow{2}{*}{SNR [dB]} & \multicolumn{2}{c}{hypothesis tested}                             \\ \cline{2-3}
                     & \multicolumn{1}{c}{\texttt{MAI}$>$\texttt{MAI}$_{RR-I}$} & \multicolumn{1}{c}{\texttt{MAI}$_{ext}>$\texttt{MAI}$_{RR-I}$} \\\specialrule{1.5pt}{0pt}{0pt}
-10 & 0.0003  & 0.0001 \\
0 & 0.0007 & 0.0046 \\
10 & 0.0023 & 0.0005 \\
\specialrule{1.5pt}{0pt}{0pt}
\multirow{2}{*}{SNR [dB]} & \multicolumn{2}{c}{hypothesis tested}                             \\ \cline{2-3}
                     & \multicolumn{1}{c}{\texttt{MPZ}$>$\texttt{MPZ}$_{RR-I}$} & \multicolumn{1}{c}{\texttt{MPZ}$_{ext}>$\texttt{MPZ}$_{RR-I}$} \\\specialrule{1.5pt}{0pt}{0pt}
-10 & 0.027 & 0.0018 \\
0 & 0.0086 & 0.0086 \\
10 & 0.0036 & 0.0007 \\
\end{tabular}
\small
\end{table}

\begin{figure}[t!] 
\centering
\includegraphics[width=0.6\linewidth]{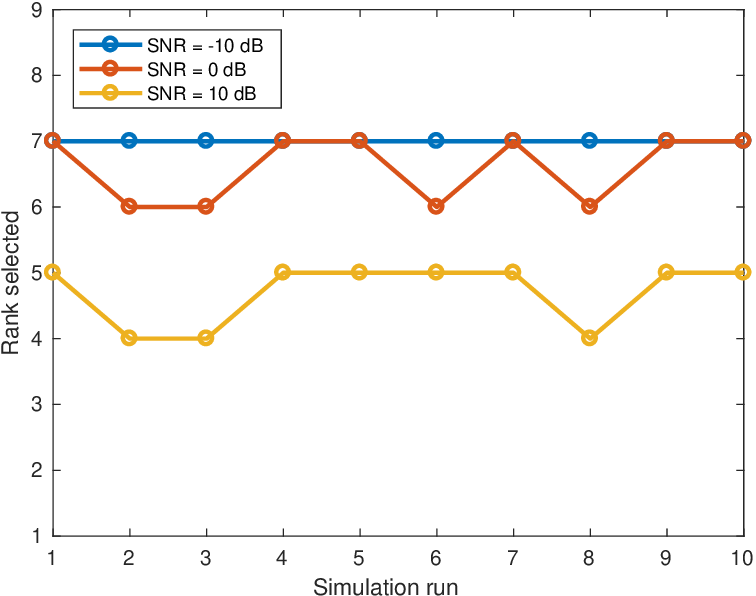}
\caption{Ranks selected across simulation runs according to criterion in (\ref{r}) for $\delta=0.8.$} \label{ranks}
\end{figure}

At each iteration $l$ of Algorithm \ref{A1}, we measured the localization error defined as the Chebyshev distance between the source found by a given activity index and the closest active source. Figs. \ref{sim1} and \ref{sim2} report averaged localization errors made by activity indices considered along with corresponding $p$-values of one-sided Mann-Whitney-Wilcoxon median test. It is seen that, in practice, SNR does matter, as the quality of estimates of $\R$ and $\N$ affects spatial resolution of indices and has an impact on the iterative implementation, \emph{cf.} also \cite{Moiseev2011}. The ranks selected for reduced-rank activity indices are shown on Fig. \ref{ranks}. It is seen that they are stable across simulation runs, and that lower ranks are actually selected for higher SNR of 10 dB. In particular, as the maximal rank selected according to criterion in (\ref{r}) for $\delta=0.8$ was 7 across all SNR and all simulation runs, we presented in Fig. \ref{sim2} the averaged localization errors for the last 2 iterations of the algorithm, i.e., when the reduced-rank approach was always active. The gain in localization error is significant in this case, especially considering that the scale on $y$-axis representing average localization error in [mm] is doubled in Fig. \ref{sim2} compared with Fig. \ref{sim1}.

\subsection{MEG ASSR Data} \label{assr}
We present a sample usage of the proposed indices implemented using Algorithm~\ref{A1} on real measurements. The data were recorded from an auditory steady state response (ASSR) experiment, where stimuli were presented to a subject as 40 Hz modulated pulses. The data were collected using a 151-channel MEG scanner (CTF/VSM) with a frequency range of
1–150 Hz and a sample rate of 600/s.

After preprocessing and artifact removal, 1181 epochs of the MEG data of 1 sec length were left for analysis. The ASSR stimuli start at 400 ms relative to the trial start. We have estimated the signal covariance matrix $\R$ using signals (concatenated across trials) from the $[400, 800]$ ms intervals, and similarly used the $[0, 400]$ ms intervals preceding stimuli to estimate the noise covariance matrix $\N.$ 

It is well known that in such experiments sustained highly correlated bilateral neural activations are observed, see \cite{Moiseev2011} and references therein. In such a case, in addition to the simplified method used in the previous section (which assumed that sources found at iterations $l>r$ are uncorrelated with all other sources), we present an estimation method for $[\Q]_{l\times l}$ for $\H'$ needed for $MAI_{RR}$ and $MPZ_{RR}$ localizers in their original formulations, \emph{cf.} (\ref{model_eq}), (\ref{RRMAI}) and (\ref{RRMPZ}). This can be achieved as follows: using notation as in Algorithm~\ref{A1}, from (\ref{SG}) in Fact \ref{relation!} in \ref{kru} it is seen that $[\Q]_{l\times l}$ can be estimated at $l$-th iteration for $i$-th source considered as 
\begin{equation} \label{latesmile}
  _i\widehat{[\Q]}_{l\times l}=\big(\H(\bth_i)^t\R^{-1}\H(\bth_i)\big)^{-1}-\\
  \big(\H(\bth_i)^t\N^{-1}\H(\bth_i)\big)^{-1},\ 1\leq l\leq l_0.
\end{equation}
We note en passant that, assuming true active sources are found at each step, such procedure leads to (up to permutation of sources' order) a gradual discovery of the true covariance matrix $\Q.$ Thus, in addition to activity indices (\ref{MAI_it})-(\ref{MPZ_RR_I_it}) introduced in the previous section in the form amenable to iterative computation using Algorithm \ref{A1}, we consider in this section also the following two versions:

\begin{itemize}
\item \texttt{MAI}$_{RR-C}$ activity index defined at $l$-th iteration as:
  \setstretch{2}
  \begin{equation}
    \texttt{MAI}_{RR-C}=\left\{
    \begin{array}{l}
      tr\{\G\S^{-1}\}-l,\ 1\leq l\leq r,\\
      tr\{\G'(\S')^{-1}\Pre^{(r)}_{\S'}\}-r,\ r<l\leq l_0.\\ 
    \end{array}\right.
  \end{equation}
  \setstretch{1}
  We note that \texttt{MAI}$_{RR-C}$=\texttt{MAI} for $1\leq l\leq r.$
\item \texttt{MPZ}$_{RR-C}$ activity index defined at $l$-th iteration as:
  \setstretch{2}
  \begin{equation}
    \texttt{MPZ}_{RR-C}=\left\{
    \begin{array}{l}
      tr\{\S\T^{-1}\}-l,\ 1\leq l\leq r,\\
      tr\{\S'(\T')^{-1}\Pre^{(r)}_{\S'}\}-r,\ r<l\leq l_0.\\ 
    \end{array}\right.
  \end{equation}
  \setstretch{1}
  We note that \texttt{MPZ}$_{RR-C}$=\texttt{MPZ} for $1\leq l\leq r.$
\end{itemize}

As in the previous section on simulated results, the solution of the forward problem has been obtained using FieldTrip toolbox \cite{FieldTrip2011}, resulting in $s=15002$ candidate positions. Specifically, we used realistic single-shell model using brain surface from subject's segmented MRI images. We used \texttt{ft\_compute\_leadfield} function and assumed that the orientation of each source is that of the lead field's principle orientation found by SVD. We did not use back-projection of weaker orientations nor lead-fields normalization to match exactly the forward model in (\ref{model}). We used the CTF coordinate system in [cm] units, with ALS orientation and the origin between the ears.

We expected the sources to be discovered primarily in auditory cortex, starting from the strongest source. In order to obtain some guidance regarding the behaviour of localizers considered, we have also evaluated the Chebyshev distance between discovered sources and the closer of the two voxels perceived to be closest to the centres of left- and right auditory areas, respectively. For brevity, we denote this distance as $d_C$ below. 

For the whole available data used (1181 trials) used for estimation of covariance matrices $\R$ and $\N$, in the first two iterations, almost all indices discovered two sources: one in the left- and one in the right auditory area, with average $d_C$ distance for the first two iterations between 1.1 cm and 1.4 cm. The only indices failing to do so were rank-1 $\texttt{MAI}_{RR-I}$ and $\texttt{MPZ}_{RR-I}$ indices, which failed to discover the second source. A likely reason is that the assumption that the first source is uncorrelated with the second one clearly does not hold in this case. 

These results align well with existing knowledge on ASSR sources which are known to be well localized and may be considered dipolar with high degree of certainty. To verify this hypothesis further, we considered the case of limited available data, with only the first 100 trials used to estimate covariance matrices $\R$ and $\N.$ We selected $l_0=4$ sources (iterations of Algorithm \ref{A1}) to discover locations of four primary sources, following the indices' saturation property observed on a similar data in \cite{Moiseev2011}. All rank constraints available $r=1,2,3$ were tested, which corresponds to setting $\delta=0.2, \delta=0.4, \delta=0.6$ in (\ref{r}), respectively. 

\begin{figure}[p!] 
\centering
\hspace*{-11cm}\includegraphics[width=2.5\linewidth]{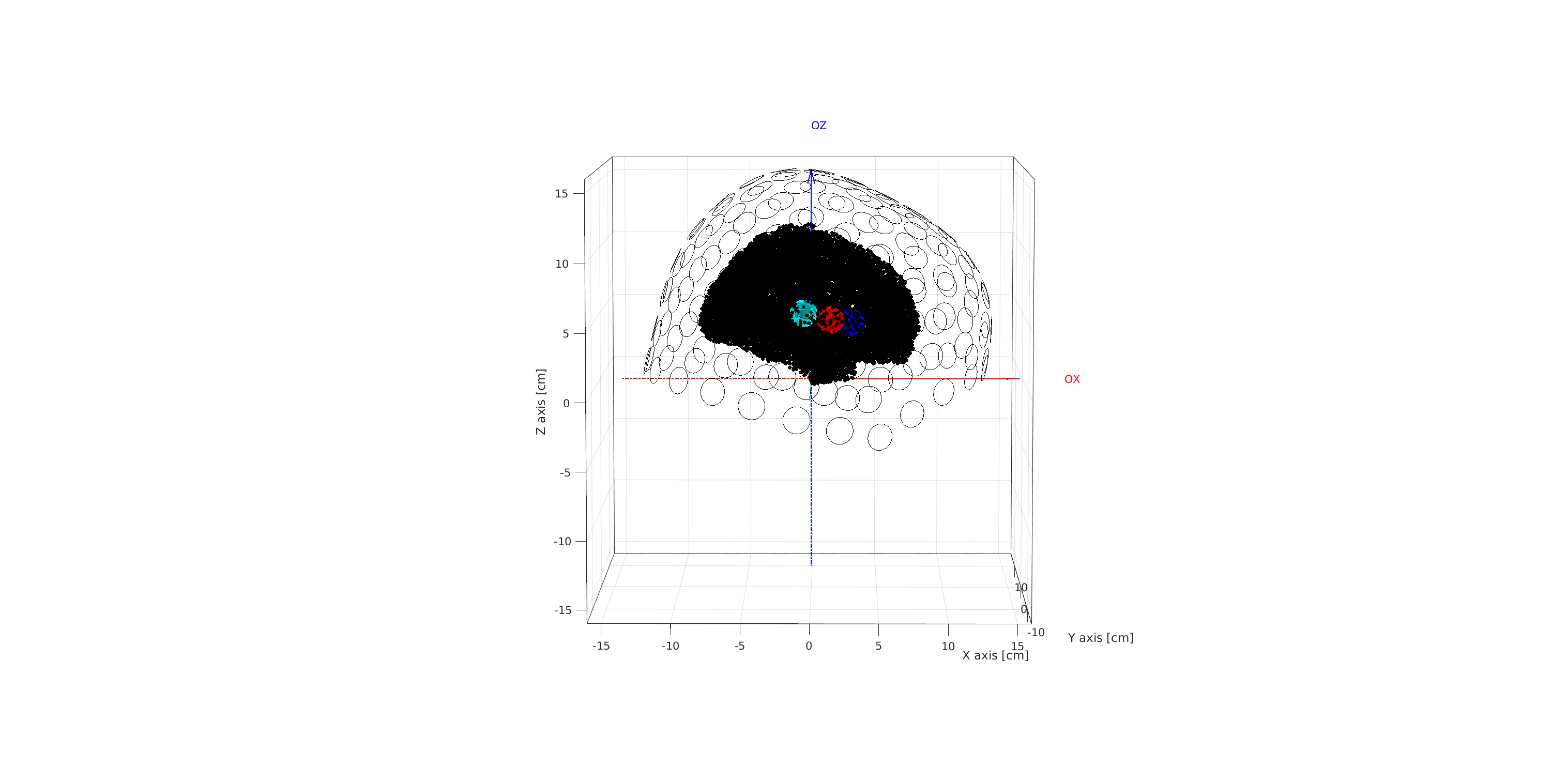}
\caption{Sources in the right auditory area discovered using available data limited to the first 100 trials with rank-1 $\texttt{MAI}_{RR-C}$ index. The blue source was discovered in the first iteration, the light blue source was discovered in the second iteration and the red source in the fourth iteration.} \label{real1}
\end{figure}

\begin{figure}[p!] 
\centering
\hspace*{-11cm}\includegraphics[width=2.5\linewidth]{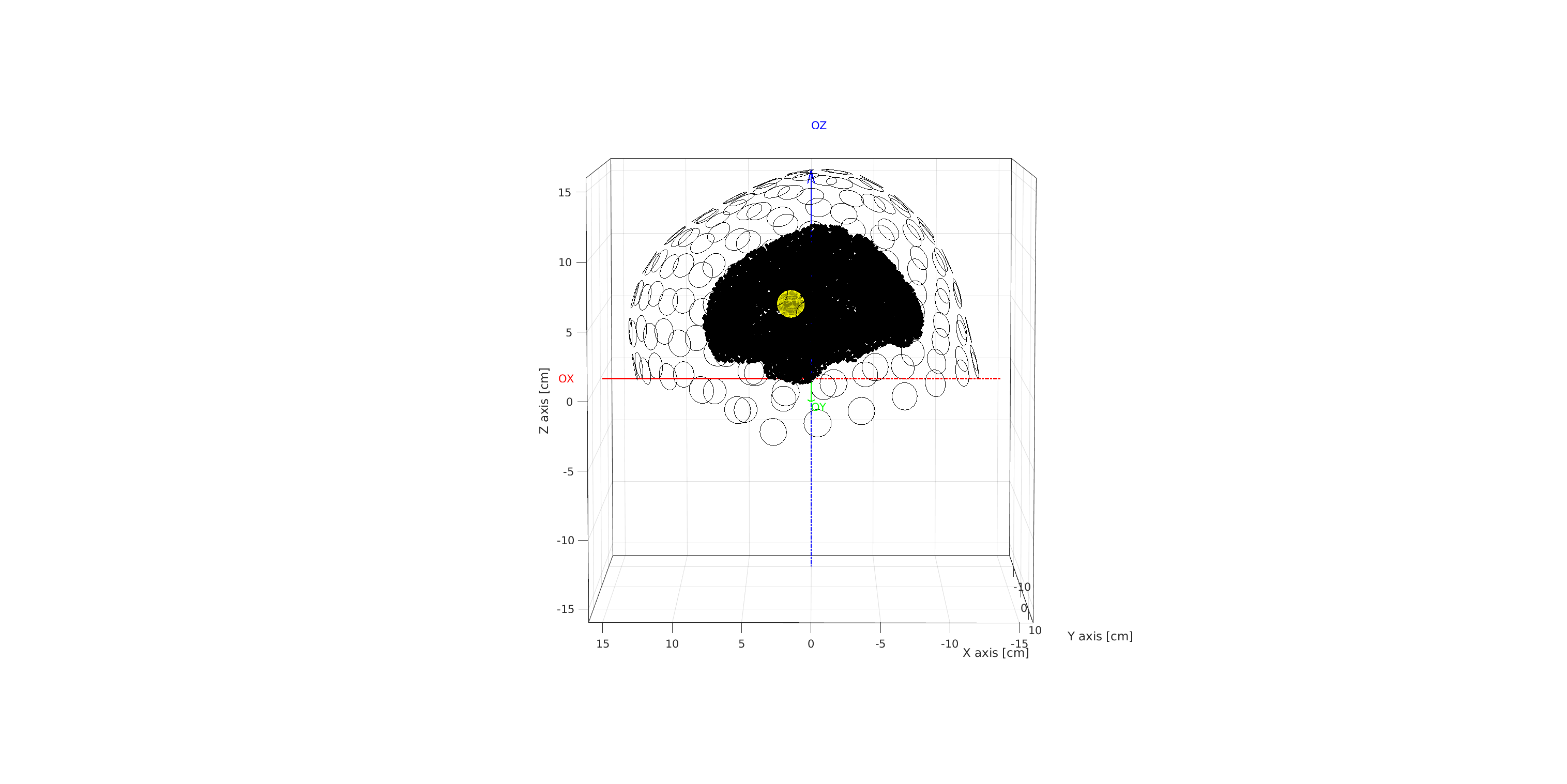}
\caption{Source in the left auditory area discovered using available data limited to the first 100 trials with rank-1 $\texttt{MAI}_{RR-C}$ index. The yellow source was discovered in the third iteration.} \label{real2}
\end{figure}

In this case, the first two iterations have discovered sources only in the right auditory area, with $d_C$ distance varying among the activity indices. The source in the left auditory area was discovered by most indices in the third iteration, \emph{cf}. examples shown in Figs. \ref{real1} and \ref{real2}. The smallest average $d_C$ distances were obtained among all rank-1 indices for the $\texttt{MAI}_{RR-C}$ index: 1.82 cm, for rank-2 for the $\texttt{MPZ}_{RR-C}$ and $\texttt{MPZ}_{ext}$ indices: 1.78 cm, and for rank-3 for the $\texttt{MAI}_{RR-C}$ and $\texttt{MAI}_{ext}$ indices: 1.97 cm. The full-rank $\texttt{MAI}$ and $\texttt{MPZ}$ indices localized sources with average $d_C$ distance of 2.43 cm from the perceived centres of left- and right auditory areas.

The above results may suggest that a stronger activity in the right auditory area was observed for these limited data. However, investigation of physiological meaning of localized sources is beyond the scope of this work. A possibility of some of them being activity indices artifacts cannot be completely excluded. Indeed, a variety of spatial filters implemented for example in \cite{Rykaczewski2020} may be used to pursue further investigation of the localized sources' activity.

We should also emphasize that the above results on MEG ASSR data, while confirming the applicability of reduced-rank indices in practical settings, should clearly not be treated as an assessment of spatial resolution of the localizers considered, as no ground truth is available for this measure on real data, which are also subject to limited MRI accuracy and co-registration errors.

\section{Conclusions}
We have introduced two novel families of reduced-rank activity indices and proved that they are unbiased for all nonzero rank constraints. We have also discussed conditions in which reducing the rank of activity index results in increased spatial resolution compared with its full-rank version. We presented a computationally efficient iterative scheme for discovering the sources on both simulated and real data. An alternative approach to source localization using proposed activity indices would be to develop a non-iterative optimization method capable of maximizing their values directly in the $l_0$-dimensional domain, which is a promising direction for our future work.

\section{Data Availability Statement}
Numerical simulations in Section \ref{nene} were prepared using open-source EEG/MEG spatial filtering framework available for download at

\noindent\url{https://github.com/nikadon/supFunSim} \cite{Rykaczewski2020}.

\noindent The data files necessary for simulations are located at

\noindent\url{http://fizyka.umk.pl/~tpiotrowski/supFunSim}.

\noindent Scripts implementing activity indices for both the numerical simulations in Section \ref{nene} and the MEG ASSR experiment described in Section \ref{assr} can be found at

\noindent\url{https://github.com/nikadon/supFunSim/tree/master/localizers}.

\noindent The MEG ASSR data files can be found at

\noindent\url{http://fizyka.umk.pl/~tpiotrowski/localizers}.

\appendix
\section{Optimization problems leading to MCMV and MV-PURE filters} \label{opt}
The MCMV filter $\W\in\mathbb{R}^{l \times m}$ is the solution of the following optimization problem \cite{Frost1972,VanVeen1997,Sekihara2008} with respect to~$\X$:
\begin{equation} \label{lcmv_opt}
\left\{
\begin{array}{ll}
\textnormal{minimize} & tr\{\X\R\X^t\}\\
\textnormal{subject to} & \X\H=\I_l,\\
\end{array}\right.
\end{equation}
where $\H\in\mathbb{R}^{m\times l}$ is the lead field matrix.\footnote{Depending on the field of application, $\H$ may be also called model or array response matrix.} We call the second condition in (\ref{lcmv_opt}) \emph{the unit-gain condition}. The solution to (\ref{lcmv_opt}) is given in (\ref{lcmv}). 

Furthermore, for a given rank constraint $r$ such that $1\leq r\leq l$, the MV-PURE filter $\W_r\in\sgen{l}{m}$ is the solution of the following problem with respect to $\X$:
\begin{equation} \label{mvp_opt}
\left\{
\begin{array}{ll}
\textnormal{minimize} & tr\{\X\R\X^t\}\\
\textnormal{subject to} & \X\in\bigcap\limits_{\iota\in\mathcal{J}}\bm{\mathcal{P}}^{\iota}_{r},\\
\end{array}\right.
\end{equation}
with $$\bm{\mathcal{P}}^\iota_r=\argmin_{\X\in\bm{\mathcal{\X}}_r^{l\times m}}\parallel \X\H-\I_l\parallel^2_\iota, \iota\in\mathcal{J},$$ and $$\bm{\mathcal{\X}}^{l\times m}_r=\{\X\in\sgen{l}{m}:rank(\X)\leq r\leq l\},$$ where $\mathcal{J}$ is the index set of all unitarily invariant norms, i.e., norms satisfying $\parallel \U\X\V\parallel_\iota=\parallel \X\parallel_\iota$ for all orthogonal $\U\in\mathbb{R}^{l \times l}$, $\V\in\mathbb{R}^{m \times m}$, and all $\X\in\mathbb{R}^{l \times m}$. The solution to (\ref{mvp_opt}) is given in \cite{Piotrowski2008} in the form (\ref{mvp}).\footnote{The work \cite{Piotrowski2008} considers a more general problem where additional linear constraints are imposed on the resulting estimator/filter. The optimization problem (\ref{mvp_opt}) and the resulting solution (\ref{mvp}) are considered in \cite{Piotrowski2008} for the case of no linear constraints.}

\begin{remark}
Comparison of optimization problems (\ref{lcmv_opt}) and (\ref{mvp_opt}) reveals that the latter is essentially a reduced-rank extension of the former, with (\ref{lcmv_opt}) being a special case of (\ref{mvp_opt}) for $r=l.$ This fact justifies calling MV-PURE filter a natural reduced-rank extension of the MCMV filter, with MV-PURE yielding minimal deviation from unit-gain constraint $\X\H=\I_l$ among filters of rank $r.$
\end{remark}  

\section{Spatial Resolution of $MAI$ and $MPZ$ Activity Indices} \label{mai_mpz_res_pd}
\begin{fact}[\cite{Moiseev2011}] \label{mai_mpz_res}
$MPZ$ has higher spatial resolution than $MAI.$
\end{fact}  
\pds The proof of the fact that $MPZ$ has higher spatial resolution than $MAI$ (\emph{cf.} Definition \ref{spatial_res_df}) has been obtained in \cite{Moiseev2011} based on inequality derived in \cite[Appendix~A]{Moiseev2011}:
\begin{equation} \label{inv}
\T\succeq \S(\G)^{-1}\S,
\end{equation}
for $\S$, $\G$, and $\T$ defined in (\ref{S}), (\ref{G}) and (\ref{T}), respectively, with equality achieved at $\bth_0.$ Namely, from (\ref{inv}) one has that
\begin{equation*}
\T^{-1}\preceq (\S(\G)^{-1}\S)^{-1}=\S^{-1}\G\S^{-1}.
\end{equation*}
Then, from the definitions of $MAI$ and $MPZ$ activity indices in (\ref{MAI}) and (\ref{MPZ}), respectively, we obtain
\begin{multline*}
  MPZ(\bth)=tr\{\S\T^{-1}\}-l=tr\{\S^{1/2}\T^{-1}\S^{1/2}\}-l\leq\\ tr\{\S^{1/2}\S^{-1}\G\S^{-1}\S^{1/2}\}-l=
  tr\{\G\S^{-1}\}-l=MAI(\bth),
\end{multline*}  
satisfying the first condition in (\ref{spatial_res}). The equality $\T(\bth_0)=\S(\bth_0)\G(\bth_0)^{-1}\S(\bth_0)$ ensures the second condition in (\ref{spatial_res}), completing the proof. 

\section{Proof of Lemma \ref{L1}} \label{L1_pd}
Let $\U\Si\V^{t}$ be the singular value decomposition of $\R^{-1/2}\H'$ with diagonal entries of $\Si$ organized in non-increasing order. Then $\H'=\R^{1/2}\U\Si\V^t$ and hence
\begin{multline} \label{bound} 
MAI_{RR}(\bth,r)+r=tr\{\V\Si^{t}\U^{t}\R^{1/2}\N^{-1}\R^{1/2}\U\Si\V^{t}\\
(\V\Si^{t}\U^{t}\U\Si\V^{t})^{-1}\V\I^{r}_{l}\V^{t}\}=\\
tr\{\Si^{t}\U^{t}\R^{1/2}\N^{-1}\R^{1/2}\U\Si(\Si^{t}\Si)^{-1}\I^{r}_{l}\}=\\
tr\{\R^{1/2}\N^{-1}\R^{1/2}\U\Si(\Si^{t}\Si)^{-1}\I^{r}_{l}\Si^{t}\U^{t}\}=\\
tr\{\R^{1/2}\N^{-1}\R^{1/2}\U\I^{r}_{m}\U^{t}\}
\leq\sum\limits_{i=1}^{r}\lambda_{i}(\R^{1/2}\N^{-1}\R^{1/2})\\
=\sum\limits_{i=1}^{r}\lambda_{i}(\R\N^{-1}).
\end{multline}
The essential inequality above is obtained from Theobald's theorem (Fact \ref{Theobald} in \ref{kru}). This proves the first inequality in (\ref{l1}). Then, to complete the proof of (\ref{l1}) we note that $\R=\H'_0(\H'_0)^t+\N$, so using (\ref{bound}) we obtain
\begin{multline} \label{eigvs}
MAI_{RR}(\bth,r)\leq\sum\limits_{i=1}^{r}\lambda_{i}(\R\N^{-1})-r=\\
\sum\limits_{i=1}^{r}\lambda_{i}(\H_{0}'(\H_{0}')^{t}\N^{-1}+\I_{m})-r=\sum\limits_{i=1}^{r}\lambda_{i}(\G'_{0}).\ 
\end{multline}

\section{Proof of Theorem \ref{Th1}} \label{Th1_pd}
\paragraph{$MAI_{RR}$ is unbiased} Remark \ref{relation!R} below Fact \ref{relation!} in \ref{kru} states that
\begin{equation} \label{s0g0}
(\S_0')^{-1}=(\G_0')^{-1}+\I_{l_0},
\end{equation}
and hence in particular
\begin{equation} \label{proc}
\Pre_{\S'_0}^{(r)}=\Pre_{\G'_0}^{(r)}.
\end{equation}
Then, by (\ref{s0g0}) and (\ref{proc})
\begin{multline} \label{unbiased!}
MAI_{RR}(\bth_{0},r)=tr\{\G_0'(\S_0')^{-1}\Pre^{(r)}_{\S_0'}\}-r=\\ tr\{(\I_{l_{0}}+\G'_{0})\Pre^{(r)}_{\G'_{0}}\}-r=
tr\{\G'_{0}\Pre^{(r)}_{\G'_{0}}\}=\sum_{i=1}^{r}\lambda_{i}(\G'_{0}),
\end{multline}
establishing unbiasedness of $MAI_{RR}$ in view of (\ref{l1}).

\paragraph{$MPZ_{RR}$ is unbiased and has higher spatial resolution than $MAI_{RR}$} From (\ref{inv}) in \ref{mai_mpz_res_pd} we obtain
\begin{equation*}
\T'\succeq \S'(\G')^{-1}\S',
\end{equation*}  
hence
\begin{equation} \label{inv2}
(\S')^{-1}\G'(\S')^{-1}\succeq (\T')^{-1},
\end{equation}
with equality achieved at $\bth_0.$ Consequently,
\begin{multline} \label{inv_eq}
\sum_{i=1}^{r}\lambda_{i}(\G'_{0})= MAI_{RR}(\bth_0,r)=tr\{\S_0'(\T_0')^{-1}\Pre^{(r)}_{\S_0'}\}-r=\\
tr\{\S_0'\big((\S_0')^{-1}\G_0'(\S_0')^{-1}\big)\Pre^{(r)}_{\S_0'}\}-r= 
MPZ_{RR}(\bth_0,r).
\end{multline}
Let now $\V\La \V^t$ be the eigenvalue decomposition of $\S'$ with eigenvalues organized in non-increasing order. Then, using similar algebraic manipulations as in (\ref{inout}) and the fact that $(\S')^{1/2}=\V\La^{1/2} \V^t$ we obtain
\begin{equation*}
\Pre_{\S'}^{(r)}\S'=\V\I^r_l\La \V^t=(\S')^{1/2}\Pre_{\S'}^{(r)}\Pre_{\S'}^{(r)}(\S')^{1/2},
\end{equation*}
and
\begin{equation*}
(\S')^{-1}\Pre_{\S'}^{(r)}=\V\La^{-1}\I^r_l\V^t=(\S')^{-1/2}\Pre_{\S'}^{(r)}\Pre_{\S'}^{(r)}(\S')^{-1/2}.
\end{equation*}
Therefore, using (\ref{inv2}) one has that
\begin{multline*}
MPZ_{RR}(\bth,r)+r=tr\{\S'(\T')^{-1}\Pre_{\S'}^{(r)}\}=\\
tr\{\Pre_{\S'}^{(r)}(\S')^{1/2}(\T')^{-1}(\S')^{1/2}\Pre_{\S'}^{(r)}\}\leq\\
tr\{\Pre_{\S'}^{(r)}(\S')^{1/2}(\S')^{-1}\G'(\S')^{-1}(\S')^{1/2}\Pre_{\S'}^{(r)}\}=\\
tr\{\G'(\S')^{-1/2}\Pre_{\S'}^{(r)}\Pre_{\S'}^{(r)}(\S')^{-1/2}\}=
tr\{\G'(\S')^{-1}\Pre_{\S'}^{(r)}\}=\\
MAI_{RR}(\bth,r)+r,
\end{multline*}
and thus $MPZ_{RR}(\bth,r)\leq MAI_{RR}(\bth,r).$ In view of unbiasedness of $MAI_{RR}$ and (\ref{inv_eq}) this completes the proof.\ 

\section{Proof of Proposition \ref{ineqs}} \label{ineqs_pd}
By applying matrix inversion lemma \cite{Horn1985} to
\begin{equation*}
\R=\H_0\Q\H_0^t+\N=\H_0'(\H_0')^t+\N
\end{equation*}
one has 
\begin{equation} \label{Woodbury}
\R^{-1}=\N^{-1}-\N^{-1}\H_0'\big(\I_{l_0}+(\H'_0)^t\N^{-1}\H_0'\big)^{-1}(\H_0')^t\N^{-1}.
\end{equation}
We obtain thus that $\G'=\S'+\Y_1$, where $\Y_1=\K'\Z(\K')^t$ with $\Z=(\I_{l_0}+\G'_0)^{-1}\succ 0$ and $\K'=(\H')^t\N^{-1}\H_0'.$ Note that the matrix $\Y_1$ is positive semidefinite, $\Y_1\succeq 0$, and let $r_1$ and $r_2$ be such that $1\leq r_1\leq r_2\leq l_0.$ Denote $\Y_p=\Pre_{\S'}^{(r_2)}-\Pre_{\S'}^{(r_1)}\succeq 0.$ Then, using similar algebraic manipulations as in (\ref{inout}) it is seen that
\begin{equation} \label{SYps}
\Y_p(\S')^{-1}\Y_p=(\S')^{-1}\Y_p\succeq 0\quad\text{and}\quad \Y_p\S'\Y_p=\Y_p\S'\succeq 0.
\end{equation}
Therefore,
\begin{multline*}
MAI_{RR}(\bth,r_2) - MAI_{RR}(\bth,r_1)=\\
tr\{\G'(\S')^{-1}[\Pre_{\S'}^{(r_2)}-\Pre_{\S'}^{(r_1)}]\}-(r_2-r_1)=\\
tr\{(\S'+\Y_1)(\S')^{-1}\Y_p\}-(r_2-r_1)=\\
tr\{\Y_1(\S')^{-1}\Y_p\}=tr\{\Y_1^{1/2}(\S')^{-1}\Y_p\Y_1^{1/2}\}\geq 0,
\end{multline*}
as $\Y_1^{1/2}(\S')^{-1}\Y_p\Y_1^{1/2}\succeq 0.$ Using now expression of $\R^{-1}$ as in (\ref{Woodbury}), we obtain that $\T'=(\H')^t\R^{-1}\N\R^{-1}\H'$ satisfies (\emph{cf.} \cite[Appendix~A]{Moiseev2011})
\begin{equation} \label{megaT}
\T'=\S'-[\K'\Z(\K')^t-\K'\Z\G'_0\Z(\K')^t].
\end{equation}
Note that $\Z^{-1}=\I_{l_0}+\G'_0\succ \G'_0$, hence $\Z\succ \Z\G'_0\Z$, and consequently $\K'\Z(\K')^t\succeq \K'\Z\G'_0\Z(\K')^t.$ From (\ref{megaT}) we obtain therefore that $\T'=\S'-\Y_2$, where $\Y_2$ is positive semidefinite, i.e., $\T'\preceq \S'$, and hence
\begin{equation} \label{TS}
(\T')^{-1}\succeq (\S')^{-1}.
\end{equation}
By definition, there exists $\Y_3\succeq 0$ such that $(\T')^{-1}=(\S')^{-1}+\Y_3.$ Using this fact, from (\ref{SYps}), and proceeding similarly as above for $MAI_{RR}$ we obtain that
\begin{multline*}
MPZ_{RR}(\bth,r_2) - MPZ_{RR}(\bth,r_1)=\\
tr\{\S'(\T')^{-1}[\Pre_{\S'}^{(r_2)}-\Pre_{\S'}^{(r_1)}]\}-(r_2-r_1)=\\
tr\{\S'[(\S')^{-1}+\Y_3]\Y_p\}-(r_2-r_1)=\\
tr\{\Y_p\S'\Y_3\}=tr\{\Y_3^{1/2}\Y_p\S'\Y_3^{1/2}\}\geq 0,
\end{multline*}
as $\Y_3^{1/2}\Y_p\S'\Y_3^{1/2}\succeq 0.$ This completes the proof.\ 

\section{Proof of Proposition \ref{superiority_ext}} \label{superiority_ext_pd}
For $i\in\{1,\dots,l\}$ one has
\begin{multline*}
\lambda_i(\G\S^{-1})=\lambda_i(\G'(\S')^{-1})=
\lambda_i\big((\G')^{1/2}(\S')^{-1}(\G')^{1/2}\big)\geq\\
\lambda_i\big((\G')^{1/2}(\S')^{-1}\Pre_{\S'}^{(r)}(\G')^{1/2}\big)=
\lambda_i(\G'(\S')^{-1}\Pre_{\S'}^{(r)}),
\end{multline*}
where the inequality follows from \cite[Corollary 7.7.4]{Horn1985}, as $(\S')^{-1}\succeq (\S')^{-1}\Pre_{\S'}^{(r)}$, \emph{cf.} also (\ref{inout}). This proves (\ref{Ext1}), as $\sum_{i=1}^r\lambda_i(\G'(\S')^{-1}\Pre_{\S'}^{(r)})=tr\{\G'(\S')^{-1}\Pre_{\S'}^{(r)}\}.$ Moreover, from Fact \ref{relation!} in \ref{kru} one has
\begin{multline} \label{powerofeq}
MAI_{ext}(\bth_0,r)=\sum_{i=1}^{r}\lambda_i(\G_0\S_0^{-1})-r=\sum_{i=1}^{r}\lambda_i(\G_0\Q)=\\\sum_{i=1}^{r}\lambda_i(\G'_0)=MAI_{RR}(\bth_0,r).
\end{multline}
The proof of (\ref{Ext2}) follows similarly as above:
\begin{multline*}
\lambda_i(\S\T^{-1})=\lambda_i(\S'(\T')^{-1})=
\lambda_i\big((\T')^{-1/2}\S'(\T')^{-1/2}\big)\geq\\
\lambda_i\big((\T')^{-1/2}\Pre_{\S'}^{(r)}\S'(\T')^{-1/2}\big)=
\lambda_i(\S'(\T')^{-1}\Pre_{\S'}^{(r)}),
\end{multline*}
where the inequality follows from \cite[Corollary 7.7.4]{Horn1985}, as in this case $\S'\succeq \Pre_{\S'}^{(r)}\S'.$ This proves (\ref{Ext2}), as $\sum_{i=1}^r\lambda_i(\S'(\T')^{-1}\Pre_{\S'}^{(r)})=tr\{\S'(\T')^{-1}\Pre_{\S'}^{(r)}\}.$ We recall now from Fact \ref{mai_mpz_res} in \ref{mai_mpz_res_pd} that equality is achieved in (\ref{inv}) for $\bth_0$, i.e., $\T_0=\S_0\G_0^{-1}\S_0$, and hence $\T_0^{-1}=\S_0^{-1}\G_0\S_0^{-1}.$ Thus,
\begin{equation*}
MPZ_{ext}(\bth_0,r)=\sum_{i=1}^{r}\lambda_i(\S_0\T_0^{-1})-r=\sum_{i=1}^{r}\lambda_i(\G_0\S_0^{-1})-r.
\end{equation*}
In view of (\ref{powerofeq}) and (\ref{th1}) in Theorem \ref{Th1} this completes the proof of (\ref{Ext3}).\ 

\section{Known results used} \label{kru}
\begin{fact}[\cite{Theobald1975}] \label{Theobald}
Let $\C\in\sgen{n}{n},\D\in\sgen{n}{n}$ be symmetric matrices. Denoting by $c_1\geq c_2\geq\dots\geq c_n$ and $d_1\geq d_2\geq\dots\geq d_n$ the eigenvalues of $\C$ and $\D$, respectively, one has 
\begin{equation} \label{poeq}
tr\{\C\D\}\leq\sum_{i=1}^n c_id_i.
\end{equation}
The equality can be attained.\footnote{The work \cite{Theobald1975} gives explicit form of eigenvalue decompositions of $\C$ and $\D$ for which equality can be attained.}
\end{fact}

The following Fact \ref{relation!} is well-known and can be obtained, e.g., by applying matrix inversion lemma \cite{Horn1985} to $\S_0.$ It is stated explicitly in \cite{Piotrowski2014b}.

\begin{fact}[\cite{Piotrowski2014b}] \label{relation!}
With notation as in model (\ref{model}), one has
\begin{equation} \label{SG}
\S_0^{-1}=\G_0^{-1}+\Q.
\end{equation}
\end{fact}  

\begin{remark} \label{relation!R}
In terms of model (\ref{model_eq}), from (\ref{SG}) we obtain in this case
\begin{equation} \label{SG_eq}
(\S'_0)^{-1}=(\G'_0)^{-1}+\I_{l_0}.
\end{equation}  
\end{remark}  

\begin{fact}[Lidskii's Theorem, p. 248 \cite{Marshall1979}] \label{Lidskiis}
If $\G,\H\in\mathbb{C}^{n\times n}$ are positive semidefinite Hermitian, and $1\leq i_1<\dots<i_k\leq n$, then
\begin{equation}
\prod_{t=1}^k\lambda_{i_t}(\G\H)\leq\prod_{t=1}^k\lambda_{i_t}(\G)\lambda_t(\H).
\end{equation}  
\end{fact} 

\section*{Acknowledgment} 
This study was supported by a grant from the Polish National Science Centre (UMO- 2013/08/W/HS6/00333). The authors would like to thank Mr. Mateusz Wilk for his help with simplifying the proof of Lemma \ref{L1} and conducting simulations. 

  %% References

  \bibliographystyle{model1-num-names}
  \bibliography{IEEEabrv,references}

\end{document}